\documentclass[onecolumn,preprint,preprintnumbers,amsmath,amssymb]{revtex4}

\usepackage{graphicx}
\usepackage{dcolumn}
\usepackage{bm}

\begin{document}
\title{Focusing of atoms with strongly confined light potentials}
\author{Lars Egil Helseth}
\address{Department of Physics, University of Oslo, P.O. Box 1048 Blindern, N-0316 Oslo, Norway}

\begin{abstract}
Focusing of atoms with light potentials is studied. In 
particular, we consider strongly confined, cylindrical symmetric potential, 
and demonstrate their applications in both red and blue-detuned focusing of 
atoms. We also study the influence of aberrations, and find that a resolution 
of 1 nm should in principle be possible.   
\end{abstract}

\maketitle

\narrowtext

\newpage

\section{Introduction}
Focusing of light is of importance both in fundamental studies and for 
technical applications. Microscopy, spectroscopy, optical data storage are 
only a few of the fields utilizing focused light. The first treatments of 
electromagnetic focusing problems are due to Ignatowsky\cite{Ignatowsky}.  However, the detailed
structure of the wave-field in the focal region was clarified by Wolf and 
coworkers, who did the first detailed studies of electromagnetic focusing 
systems\cite{Wolf,Richards,Boivin}. Here the socalled Debye 
approximation was adopted, where only the plane waves with propagation vectors 
that fall inside the geometrical cone whose apex is at the focal point 
contribute to the field in the focal region\cite{Wolf,Debye}. The Debye
approximation predicts that the electromagnetic field is symmetrical about the
focal plane. However, later it was found that the Debye theory is valid only when the focal point is
located many wavelengths away from the aperture, and the angular aperture is
sufficiently large\cite{Wolf1}. When these conditions are not fulfilled, 
the electromagnetic field is not symmetrical about the focal 
plane\cite{Stamnes,Stamnes1,Li,Dhayalan}.

Recently, focusing of spatially inhomogenous polarization distributions have triggered a
lot of interest due to their potential applications in data storage and for
manipulation of molecules\cite{Helseth,Youngworth,Biss,Novotny}. Such 
distributions may be efficiently produced by spatial light modulators (e.g. LCD's) or
interferometric methods\cite{Tidwell,Tidwell1,Grosjean,Bomzon}. The 
potential of of these polarization distributions is not yet fully exploited, 
and many interesting discoveries may occur in the future. Of particular
interest here is effective focusing of atoms.  
   
Focusing of neutral atoms using light has attracted considerable interest the last 25
years, since the atomic wavelength is much smaller than the optical wavelength,
and because atoms interact with matter in a different manner than other
particles. An atom in a radiation field experiences two different
forces\cite{Ashkin,Minogin1}. The spontaneous 
emission force results from absorption and random spontaneous emission of
photons, and is basically radiation pressure. The magnitude of this force is
limited by the rate of spontaneous emission, and saturates as the laser
intensity increases. The second force is the socalled gradient or dipole force, which
derives from the interaction of the induced atomic dipole moment with the
nonuniform light distribution. It can be made arbitrary large by
increasing the intensity gradient, and is dependent on the amount of detuning
from the atomic resonance frequency. For red-detuned light the force is in the
direction of increasing intensity, whereas for blue-detuned light the sign of
the force is reversed. The first demonstration of focusing of an atomic beam
with the dipole force was presented by Bjorkholm et al. 
\cite{Bjorkholm,Bjorkholm1}. They showed that using a red-detuned 
continous wave laser a spot size of 28 $\mu m$ is obtainable. The
resolution was here limited by spontaneous emission processes, which
effectively introduced aberrations in the focused beam. Later Balykin et al.       
reported experiments using two counterpropagating laser beams acting as an
atom lens\cite{Balykin,Balykin1}. In this way they were able to demonstrate imaging of atomic
sources, but were limited by similar aberrations as
Bjorkholm et al.\cite{Bjorkholm,Bjorkholm1}. To avoid this, Balykin and
Letokhov suggested the application of a hollow, blue-detuned laser 
beam. This dipole potential has the advantage that the atoms go through a
region of low intensity, so that the spontaneous emission is rather small. 
Two more detailed studies of this type of lens were later presented in
Refs. \cite{Gallatin,McClelland}. 

More recently, focusing of atoms using a standing wave field has attracted
considerable 
attention\cite{McClelland1,Natarajan,Meneghini,Anderson,Cohen,Lee1,Prentiss,Sleator,Timp}. 
When an atomic beam is incident transverse to a standing wave field, the field
gradient acts as a sequence of lenses for the atoms. At the focal plane the
atomic spatial distribution consists of a periodic set of lines or dots
distanced from each other by $\lambda /2$, where $\lambda$ is the wavelength of
the light. Atomic beams of Na, metastable He and Cr has been focused 
using standing wave fields\cite{McClelland1,Prentiss,Sleator,Timp}. With Cr 
widths as small as a few tens of nanometers have been produced\cite{Anderson}. 
However, it has been pointed out that its application in atomic lithography is
restricted, since arbitrary patterns cannot be produced. Thus, some 
alternative methods have been suggested. For example, it has been proposed 
that it is possible to form an atom lens using the near field from a
fiber tip, thus obtaining resolutions of less than 10 nm\cite{Balykin2}. 
A particularly interesting proposal was given by Dubetsky and
Berman\cite{Dubetsky}. They proposed to use a conical mirror  
and an inhomogeneously polarized pulsed laser beam to obtain the wanted dipole 
potential. This configuration is in many ways similar to that of 
Bjorkholm et al., but avoids the problems with spontaneous
emission aberrations.    

In the current paper we investigate how a strongly confined light potential can
be used to focus atoms. We are particularly interested in beams with cylindrical
symmetry, since these provide a symmetric optical potential. To
that end, we first need to determine the properties of such light potentials.
Here we use the Debye approximation to calculate the electric field in the focal
region. In particular, it is found that a radially polarized beam may be
used in red-detuned focusing, whereas the azimuthal distribution is useful 
for blue-detuned focusing. 
   
\section{High aperture focusing of light}
In previous studies only weakly focused light was
considered\cite{Balykin,Gallatin}. However, it is also of interest to 
consider strongly confined light potentials generated by high aperture
focusing. Such potentials may give higher intensity gradients and shorter
atomic focal lengths, which is of interest in certain applications. 
However, if linearly polarized light is strongly focused, the light intensity 
is no longer symmetric in the focal region, which may reduce the resolution of 
an atom focusing system. Thus, it is necessary to examine the properties of
strongly focused light more carefully. We start by assuming that our focusing system has high 
angular aperture and that the focal point is placed many wavelengths away from 
the aperture. Then the diffracted field near the focal plane can be calculated 
in the Debye approximation as\cite{Stamnes} 

\begin{equation}
\mbox{\boldmath $E$} =-\frac{ikC}{2\pi}\int \int_{\Omega}
\mbox{\boldmath $T$} (\mbox{\boldmath $s$})
exp[ik(s_{x}x+s_{y}y+s_{z}z)] ds_{x}ds_{y} \,\,\, ,
\label{a}
\end{equation}
where C is a real constant, $k=2\pi /\lambda $ is the wavenumber, 
$\mbox{\boldmath$s$}=(s_{x},s_{y},s_{z})$ is the unit vector along a 
typical ray, $\Omega$ is the solid angle formed by all the geometrical 
rays, $\mbox{\boldmath$T$}(\mbox{\boldmath $s$})$ is the vector pupil 
distribution which accounts for the polarization, phase and amplitude 
distributions at the exit pupil. 

In spherical coordinates, the unit wave vector is defined as
\begin{equation}
\mbox{\boldmath $s$}=(sin\theta cos\phi, sin\theta sin\phi, cos\theta )\,\,\, .
\label{d}
\end{equation}
The position vector can be written as
\begin{equation}
\mbox{\boldmath $r$}_{c}=(r_{c}sin\theta_{c} cos\phi_{c}, r_{c}sin\theta_{c}
sin\phi_{c}, z)  \,\,\, .
\label{e}
\end{equation}
This gives the following diffraction integral:
\begin{equation}
\mbox{\boldmath $E$} =-\frac{iC}{\lambda}\int_{0}^{\alpha} \int_{0}^{2\pi} \mbox{\boldmath
$T$}(\theta , \phi) \exp[ik \kappa ] sin\theta d\theta d\phi \,\,\, ,
\label{f}
\end{equation}
where $\alpha$ is the convergence semiangle and
\begin{equation} 
\kappa =zcos\theta  +r_{c} sin\theta sin\theta_{c} cos(\phi -\phi_{c}) \,\,\, .
\label{h}
\end{equation}
The amplitude vector for each ray is given by
\begin{equation}
\mbox{\boldmath $T$}(\theta ,\phi) = \mbox{\boldmath $P$}(\theta ,\phi) B(\theta
,\phi) \,\,\,  ,
\label{g}
\end{equation}
where $\mbox{\boldmath $P$}(\theta ,\phi)$ is the polarization distribution and
$B(\theta ,\phi)$ the amplitude and phase distribution at the exit pupil. We emphasize that 
$ B(\theta,\phi)$ can be used to desribe any focusing systems (with lenses and 
mirrors). For an aplanatic system $B(\theta) =
\sqrt{cos\theta}$. Other examples are found in Ref. \cite{Stamnes}. 

\begin{figure}
\includegraphics[width=12cm]{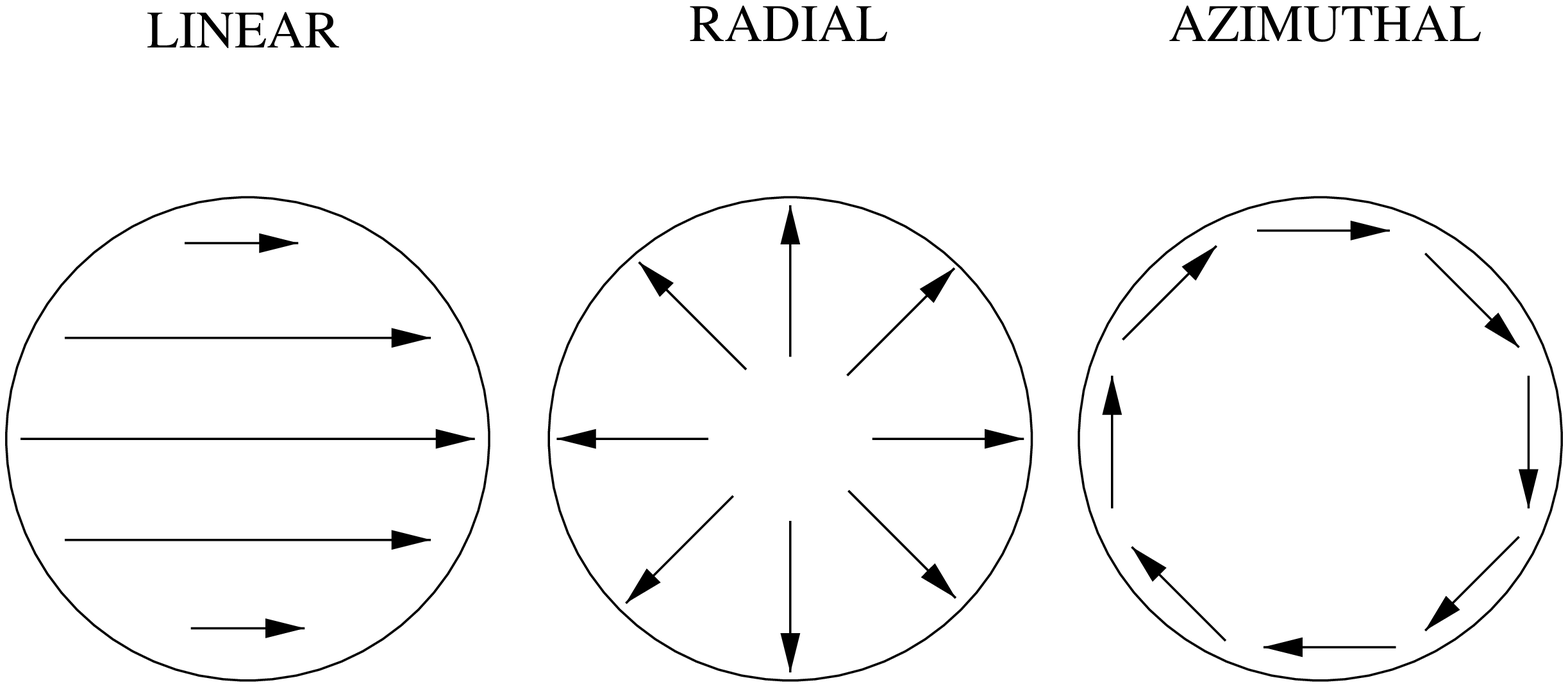}
\caption{\label{f1} The three different polarization modes considered in this paper.}
\vspace{2cm}
\end{figure}

The state of the polarization incident on the focusing system influences
the structure of the elctric field near the focal plane. To discuss this 
question quantitatively, we must find a general expression for the polarization vector. To this end, we 
assume a polarization which may in general depend on 
the polar and azimuthal angles
\begin{displaymath}
\mathbf{P_{0}} = 
\left[ \begin{array}{ccc} 
a(\theta, \phi) \\
b(\theta, \phi) \\
0                   \\ 
\end{array} \right] \,\,\, .
\end{displaymath}
Then the polarization vector can be written as
\begin{equation}
\mathbf{P}(\theta ,\phi)=R^{-1}CRP_{0} \,\,\, .
\end{equation}
Here $\mbox{\boldmath $R$}$ describes the rotation of the coordinate system 
around the optical axis
\begin{displaymath}
\mathbf{R} = 
\left[ \begin{array}{ccc} 
cos\phi & sin\phi & 0 \\
-sin\phi & cos\phi & 0 \\
   0       & 0        & 1 \\
\end{array} \right] \,\,\,  ,
\end{displaymath}
and $\mbox{\boldmath $C$}$ describes the change of polarization on propagation 
through the lens
\begin{displaymath}
\mathbf{C} = 
\left[ \begin{array}{ccc} 
cos\theta & 0   &sin\theta  \\
0               & 1        & 0 \\
-sin\theta &0    & cos\theta \\
\end{array} \right] \,\,\, .
\end{displaymath}
The result is
\begin{displaymath}
\mathbf{P}(\theta,\phi) = 
\left[ \begin{array}{ccc} 
a[cos\theta cos^{2}\phi +sin^{2}\phi ]
+b[cos\theta sin\phi cos\phi - sin\phi cos\phi ] \\
a[cos\theta cos\phi sin\phi - sin\phi cos\phi ] +
b[cos\theta sin^{2}\phi +cos^{2}\phi ] \\
-asin\theta cos\phi - bsin\theta sin\phi                \\ 
\end{array} \right] \,\,\, .
\end{displaymath}
The procedure presented above can be used to calculate the electric field in the
focal region for any combinations of polarization, phase and amplitude
distributions at the exit pupil. Here we will consider only two different polarization distributions;  
radial and azimuthal polarized light, see Fig. \ref{f1}. In both
cases a singularity occurs at the origin. Thus, practical generation of these 
beams usually involves blocking out the central portion of the beam. However, 
in our theoretical calculations the singularity does not affect the results, 
and we may ignore it (but keeping in mind its existence).  

\subsection{Radial polarization}
In the case of radial polarization we may write $a(\phi)=cos\phi$ and 
$b(\phi)=sin\phi$, which results in the following expression:

\begin{displaymath}
\mathbf{P}(\theta ,\phi) = 
\left[ \begin{array}{ccc} 
cos\theta cos\phi  \\
cos\theta sin\phi \\
-sin\theta  \\ 
\end{array} \right] \,\,\, .
\end{displaymath}
When inserted in Eq. (\ref{f}), this gives
\begin{equation}
E_{x}^{rad}=\frac{2C\pi}{\lambda}I_{1}^{rad}cos\phi_{c} \,\,\, ,
\label{j}
\end{equation}
\begin{equation}
E_{y}^{rad}=\frac{2C\pi}{\lambda}I_{1}^{rad}sin\phi_{c} \,\,\, ,
\label{k}
\end{equation}
\begin{equation}
E_{z}^{rad}=-\frac{2C\pi i}{\lambda}I_{0}^{rad} \,\,\, ,
\label{l}
\end{equation}
where
\begin{equation}
I_{0}^{rad}=\int_{0}^{\alpha } B(\theta ) sin^{2}\theta
J_{0}(kr_{c}sin\theta sin\theta_{c}) \exp(ikzcos\theta) d\theta \,\,\, ,
\end{equation}
and
\begin{equation}
I_{1}^{rad}=\int_{0}^{\alpha } B(\theta ) cos\theta 
sin\theta
J_{1}(kr_{c}sin\theta sin\theta_{c} ) \exp(ikzcos\theta ) d\theta \,\,\, .
\end{equation}  

\begin{figure}
\includegraphics[width=12cm]{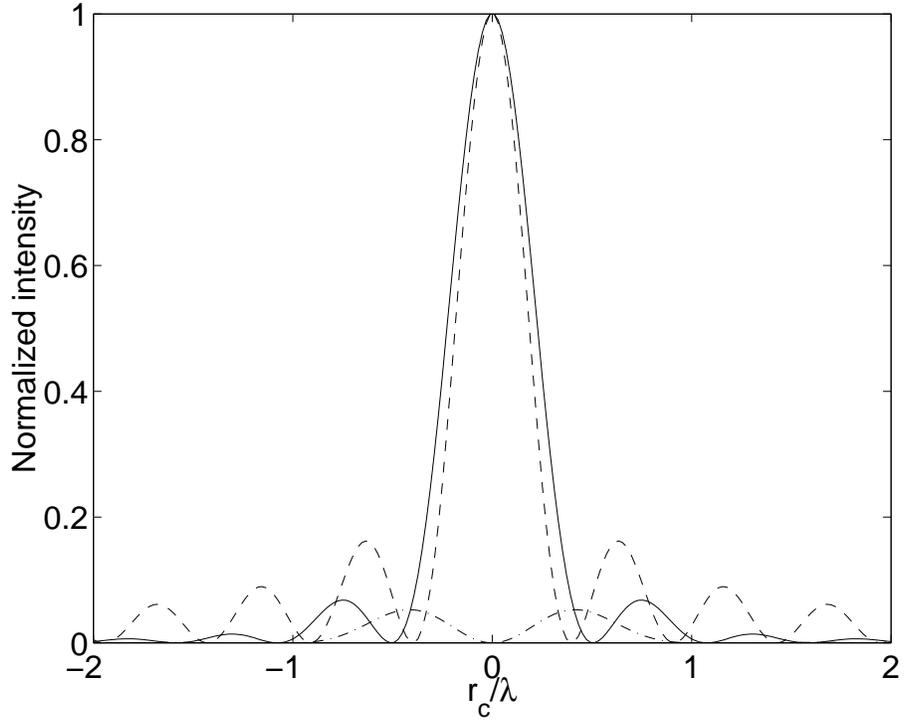}
\caption{\label{f2} $|E_{z}|^{2}$ (solid line) and 
$|E_{x}|^{2}$ (dash-dotted line) at the geometric focal point when 
$\alpha=80^{\circ}$ and $\phi _{c}=0^{\circ}$. The dashed line shows the case 
when only a narrow annular aperture is used (see text).}
\vspace{2cm}
\end{figure}

Here the z component is completely independent of the azimuthal angle, and its
importance increases with increasing angular aperture. In Fig. \ref{f2} $|E_{z}|^{2}$ (solid
line) and $|E_{x}|^{2}$ (dash - dotted line) are shown for the case 
$B(\theta) = \sqrt{cos\theta}$, $\alpha =80 ^{\circ}$ and 
$\phi _{c} =0 ^{\circ}$. Here the peak of $|E_{x}|^{2}$ is only 5 $\%$ of
$|E_{z}|^{2}$. The x component can be made much smaller by using an aperture 
wich blocks out the central part of the beam. The dashed line in Fig. \ref{f2} 
shows the case when $\alpha _{1} =70 ^{\circ}$ has been blocked out and 
and the total semiconvergence angle is $\alpha =80 ^{\circ}$. Note
that although the central lobe has shrinked, more energy is distributed into
the sidelobes. This is a general feature valid not only for radially
polarized light (see e.g. Ref. \cite{Stamnes} and references therein). 

Let us now consider an annular aperture with $\alpha  \sim 90 ^{\circ}$, B=1 and
where most of the central part (except the rim) of the exit pupil is blocked out. 
Then the x and y components of the electric
field at the focal plane are almost zero. On the other hand, for z=0 the 
z component can be approximated by 
\begin{equation}
E_{z}^{rad}\approx -\frac{2C_{1}\pi i}{\lambda} J_{0}(kr_{c}) \,\,\,  ,
\label{radpol2}
\end{equation}
where $C_{1}$ is a constant. 
We assume here that all the light is distributed into the transmitting zone 
in order to optimize the light intensity. This can be done by 
using a holographic beamshaper or an axicon combined with a proper lens
system\cite{Lee}. Close to the geometric focal point the intensity becomes
\begin{equation}
I^{r}=\frac{1}{2\eta} |E_{z}^{rad}|^{2} \approx \frac{2}{\eta}
\left(\frac{C_{1}\pi }{\lambda} \right)^{2} \left(1- \frac{(kr_{c})^{2}}{2}
\right) \,\,\,  ,
\label{radpol3}
\end{equation}
where $\eta = 377$ $\Omega$ is the impedance of vacuum.
We see that near the maximum the intensity behaves as a parabola, which
may be useful for focusing of atoms.

\subsection{Azimuthal polarization}
\begin{figure}
\includegraphics[width=12cm]{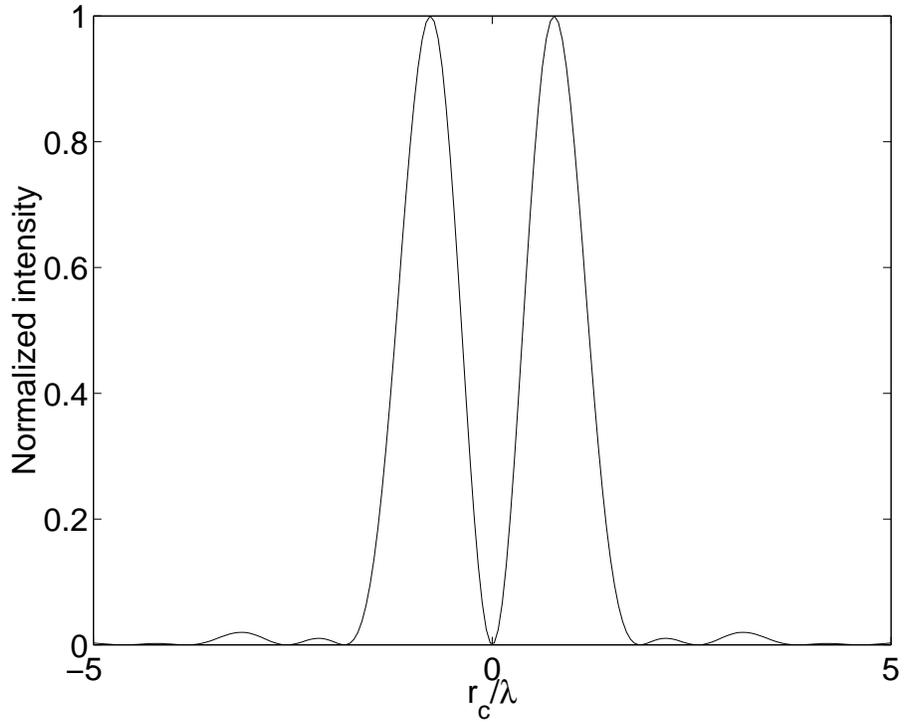}
\caption{\label{f3} $|E|^{2}$ at the focal point when $\alpha$ = $30^{\circ}$. }
\vspace{2cm}
\end{figure}

Another polarization mode of interest is the azimuthal
distribution. In this case we may write $a=sin\phi$ and 
$b=-cos\phi$, which results in  
\begin{displaymath}
\mathbf{P}(\theta ,\phi) = 
\left[ \begin{array}{ccc} 
sin\phi  \\
-cos\phi \\
0  \\ 
\end{array} \right] \,\,\, .
\end{displaymath}
When inserted in Eq. (\ref{f}), we find:
\begin{equation}
E_{x}^{asi}=\frac{2C\pi}{\lambda} I_{1}^{asi}sin\phi_{c} \,\,\, ,
\end{equation}
\begin{equation}
E_{y}^{asi}=-\frac{2C\pi}{\lambda}I_{1}^{asi}cos\phi_{c} \,\,\, ,
\end{equation}
\begin{equation}
E_{z}^{asi}=0 \,\,\, , 
\end{equation}
where
\begin{equation}
I_{1}^{asi}=\int_{0}^{\alpha } B(\theta) sin\theta
J_{1}(kr_{c}sin\theta sin\theta_{c} ) \exp(ikzcos\theta ) d\theta \,\,\, .
\end{equation}
As expected, the electric field near focus has no longitudinal component. 
Furthermore, we note that electric field vector is circularly symmetric, and is 
zero on the optical axis. In Fig. \ref{f3} the solid line shows the 
intensity  $|E|^{2}$ at the focal plane for the case $B(\theta) =
\sqrt{cos\theta}$ and $\alpha = 30^{\circ}$. Note that the two peaks are 
located approximately $\lambda$ away from the origin, and this distance is 
tuned by altering the angular aperture.  

Let us now assume an annular aperture where 
B=1, $\alpha \sim 90 ^{\circ}$ and most of the exit pupil (except the
rim) is blocked out. Then we may write (for z=0)
\begin{equation}
I^{a} \approx \frac{2\pi ^{2} C_{1}^{2}}{\lambda ^{2} \eta} J_{1}(kr_{c}) \,\,\,
. 
\label{asa}
\end{equation}
For small distances from the geometric focal point, the intensity can be written as
\begin{equation}
I^{a} \approx\frac{1}{2} \left(\frac{C_{1}k\pi r_{c}}{\lambda}\right)^{2} \,\,\,
.
\label{as1}
\end{equation}
As for the radial polarization, the intensity near the focal point behaves as 
a parabola, but now the origin is at the minimum intensity.

\section{Atom focusing} 
To calculate the the evolution of the atomic beam we use the path-integral
technique of Refs. \cite{Gallatin,Feynman}. Here the wave function
is given by
\begin{equation}
\Psi (\mbox{\boldmath $r$}_{b},t_{b}) =\int d^{3} \mbox{\boldmath $r$}_{a}
K(\mbox{\boldmath $r$}_{b},t_{b},\mbox{\boldmath $r$}_{a},t_{a})\Psi
(\mbox{\boldmath $r$}_{a},t_{a}) \,\,\, ,
\end{equation}   
where $\Psi (\mbox{\boldmath $r$}_{a},t_{a})$ is the initial wavefunction and 
$K(\mbox{\boldmath $r$}_{b},t_{b},\mbox{\boldmath $r$}_{a},t_{a})$ is the 
propagator 
\begin{equation}
K(\mbox{\boldmath $r$}_{b},t_{b},\mbox{\boldmath $r$}_{a},t_{a}) =\int_{a}^{b}
\delta \mbox{\boldmath $r$} exp\left[ \left\{ \frac{i}{\hbar} S[\mbox{\boldmath
$r$} (t)] \right\} \right] \,\,\, .
\end{equation} 
The integration is over all possible paths $\mbox{\boldmath $r$} (t)$ that
starts at ($\mbox{\boldmath $r$} _{a}, t_{a}$) and ends at 
($\mbox{\boldmath $r$} _{b}, t_{b}$), and S denotes the action evaluated along
that path:
\begin{equation}
S[\mbox{\boldmath $r$} (t)] =\int_{t_{a}}^{t_{b}}
dt\left\{ \frac{m}{2}[\partial _{t}\mbox{\boldmath
$r$} (t) ]^{2} -U(\mbox{\boldmath $r$}) \right\} \,\,\, .
\end{equation}

In general, the spatial variations of the optical potential are slow compared
to the atomic wavelength, which means that the propagator integral can be
evaluated by the method of stationary phase, thus resulting in\cite{Gallatin,Feynman} 
\begin{equation}
K(\mbox{\boldmath $r$}_{b},t_{b},\mbox{\boldmath $r$}_{a},t_{a}) \approx N 
\exp\left\{ \frac{i}{\hbar} S_{c}[\mbox{\boldmath $r$} (t)] \right\} \,\,\, .
\end{equation} 
This approximation is a semiclassical one, since 
$S_{c}[\mbox{\boldmath $r$} (t)]$ is now evaluated along the classical 
trajectory. N is in general a time and space dependent normalization factor which
varies slowly compared with the phase factor, and is therefore neglected here.
We now consider a collimated monoenergetic atomic beam travelling
parallel to the optical axis with velocity $\mbox{\boldmath $V$} _{1}$. 
Well defined collimation of Cr was demonstrated by Scholten $et$
$al$.\cite{Scholten}, and is therefore within the reach of current technology. 
The initial wave function can then be taken as
\begin{equation}
\Psi (\mbox{\boldmath $r$}_{a},t_{a}) =\exp[ikz_{a} -(i/\hbar )Et_{a} ] \,\,\, .
\end{equation}

Ovchinnikov has pointed out the importance of using 
pulsed lasers for focusing of atoms in order to minimize spontaneous decay 
aberrations\cite{Ovchinnikov}. We will therefore assume that the potential is
switched on and off so fast that the atomic kinetic energy does not change
during the switching time, i.e. $\tau\ll 1/\omega $, where $\omega =\hbar k^{2}
/2m$ is the recoil frequency of an atom with mass m.
At the time $t = -\tau /2$ an optical pulse of duration $\tau$ is applied, 
thus resulting in an attraction or repulsion of the atoms. In fact, the 
intensity gradient of the laser beam results in a force $\mbox{\boldmath $F$}
=-\mbox{\boldmath $\nabla$} U(\mbox{\boldmath $r$}) $, where 
\begin{equation}
U(\mbox{\boldmath $r$}) =\frac{\hbar \Gamma ^{2}}{8\Delta} \frac{I}{I_{s}}  \,\,\, .
\label{dipole1}
\end{equation}
Here I is the laser intensity, $I_{s}$ is the atomic transition saturation
intensity, $\Gamma$ is the natural linewidth, and $\Delta =\omega _{L} -\omega
_{R}$ is the detuning of the laser frequency $\omega _{L}$ from the atomic
resonance frequency $\omega _{R}$. For this potential to be valid the laser must 
be detuned far from the transition, $|\Delta| \gg \Gamma$.

\begin{figure}
\includegraphics[width=12cm]{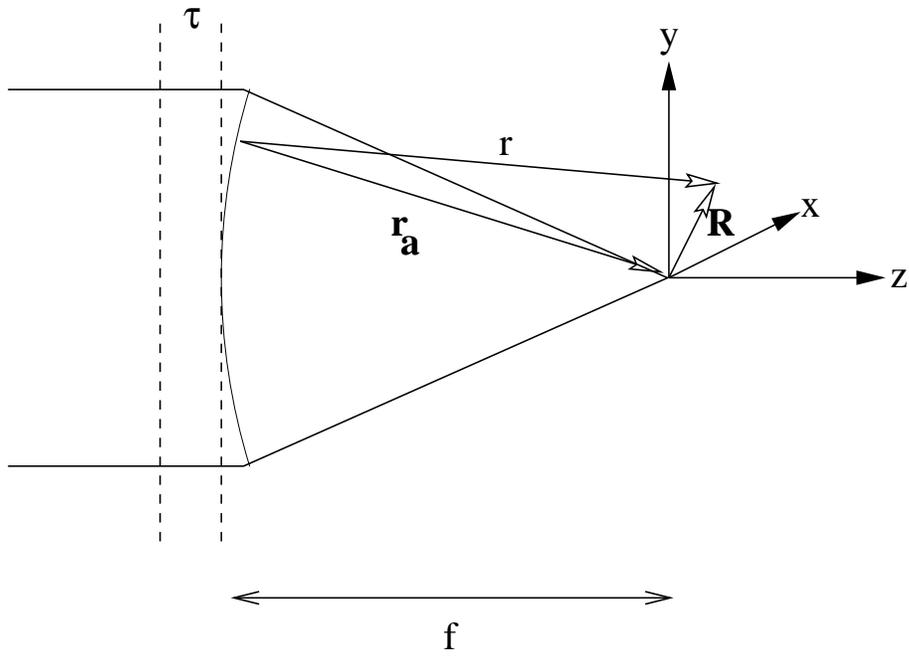}
\caption{\label{f4} Focusing of a collimated atomic beam.}
\vspace{2cm}
\end{figure}

After the optical potential is turned off, the atoms 
again behave as free particles travelling with constant velocity, see Fig.
\ref{f4}. In fact, their wavefunctions represent a spherical wavefront converging to the focal region, where the 
aberrations depend on the detailed shape of the optical potential. 
The atoms move with a velocity given by  
\begin{equation}
\mbox{\boldmath $V$} _{2} =\frac{\mbox{\boldmath
$r$}_{a}  +\mbox{\boldmath $R$}}{t} \,\,\, .
\end{equation} 
Here $\mbox{\boldmath $r$}_{a}$=$(x_{a},y_{a},z_{a})$ is the position vector from the
aperture to the 'focal point', 
$\mbox{\boldmath $R$}$=$(x,y,z)$ is the position vector from the origin to the
observation point, and $t$ the time of flight. For a perfect optical potential,
the focal distance is given by $f=V_{2}t$.

We can write the classical action as
\begin{equation}
S_{c}\left[ \mbox{\boldmath $r$} (t) \right] =\frac{m \left( \mbox{\boldmath
$r$}_{a} +\mbox{\boldmath
$R$} \right)^{2} }{2t} - U(\mbox{\boldmath $r$})\tau \,\,\, .
\end{equation}
Since we work with small angular apertures and assumes that $R \ll r_{a}$, the time of flight is nearly equal for all
atoms, i.e. independent of position. Moreover, $V_{2} \approx V_{1}$.
If we consider only those atoms located near the focus of the optical 
potential, we can omit the integration over $z_{a}$, and the wavefunction in 
the focal region becomes 
\begin{equation}
\Psi (x,y) \propto \int \int
dx_{a}dy_{a} \exp \left[ \frac{m\mbox{\boldmath $r$}_{a} \mbox{\boldmath 
$\cdot$ \mbox{\boldmath $R$}} }{\hbar t} +i\Phi(r_{a}) \right] \,\,\, ,
\label{wavecoll}
\end{equation} 
where 
\begin{equation}
\Phi(r_{a}) = \frac{m(x_{a} ^{2} + y_{a} ^{2} )}{2\hbar t}- \frac{U(\mbox{\boldmath $r$} _{a})\tau }{\hbar}
\,\,\, .\end{equation}
Here the integration is taken over the aperture set up by the laser pulse.
We see that the if the optical potential is harmonic, the quadratic
phase-factor vanishes. This equation is similar to that of the 
electric field amplitude obtained in the scalar Debye approximation. Thus 
we may say that Eq. (\ref{wavecoll}) is the Debye approximation for focusing 
of atoms.

Let us consider a circular aperture for the atomic beam. Then we may switch to cylindrical 
coordinates, and set
\begin{equation}
\mbox{\boldmath $r$}_{a} = (\rho cos\gamma, \rho sin\gamma, f - \frac{\rho
^{2}}{2f} ) \,\,\, ,
\end{equation}
and
\begin{equation}
\mbox{\boldmath $R$} = (r cos\beta, r sin\beta, z ) \,\,\, ,
\end{equation}
where we have used the expansion $ z_{a}=\sqrt{f^{2} - \rho ^{2}} \approx f - \frac{\rho ^{2}}{2f} $,
which is valid for reasonably small angular apertures.  
The wavefunction in the focal region is then expressed by
\begin{equation}
\Psi (r) \propto \int _{0}^{a}
J_{0} \left( \frac{2\pi \rho r} {\lambda _{D} f} \right) \exp\left( -i\frac{\pi \rho
^{2}}{\lambda _{D}f^{2}} z \right) \exp \left[i\Phi(\rho ) \right] \rho d\rho \,\,\, .
\label{D}
\end{equation} 
where a is the aperture radius and
\begin{equation}
\Phi(\rho) = \frac{\pi \rho ^{2}}{\lambda _{D} f}- \frac{U(\rho )\tau }{\hbar}
\end{equation}
To derive the wavefunction for this particular symmetry we have used the following identity: 
\begin{equation}
\int_{0}^{2\pi} \exp(ik\rho cos\phi )d\phi =2\pi J_{0} (k\rho) \,\,\, .
\end{equation}

Equation (\ref{D}) is similar to that given in Ref.\cite{Born}, but with the atomic 
de Broglie wavelength and a different aberration
function. The atomic density can now be found by computing $|\Psi (r)|^{2}$.
Although the similarities between focusing of light and atoms are well
known\cite{Bjorkholm,Balykin1}, we have not seen Eq. (\ref{D}) been derived 
from a path-integral approach in any previous papers. Note in particular that 
Eq. (\ref{D}) can be used to compute the wavefunction slightly away from focus for any potential of optic 
or magnetic origin, although this property will not be used here. On the other hand, it is also important to point out that this approximation is not 
valid for large convergence angles or large wavefront aberrations, as is also
the case for the scalar Debye approximation for focusing of light\cite{Born}
(Eq. (\ref{f}) is the electromagnetic Debye approximation, which is valid for
large convergence angles). 
Furthermore, the path-integral approach clearly does not take 
into account any polarization effects associated with e.g. permanently induced 
dipoles. That is, we assume that the only effect of the focused light is to 
give the atoms a 'kick' in the right direction. Finally, we have ignored 
the small effects which may be induced by the quantized electromagnetic 
field\cite{Rohwedder}. 

\subsection{Red-detuned focusing}
A red-detuned gradient potential utilizes a large and negative atom-field
detuning, which means that the atom will be attracted to
the high intensity region. To obtain a symmetrical wavefunction in the focal
region, it is of importance that the gradient potential also is symmetric. To
that end, both linear and radial polarization distributions may be used, but 
with certain restrictions. For example, focusing of linearly 
polarized light gives rise to both y and z components of the focused light 
which are proportional to $sin2\phi _{c}$ and $cos\phi _{c}$, respectively. 
The z component may have a peak intensity of 25 $\%$ compared 
to the x component, and could therefore alter the gradient
potential\cite{Helseth}. However, these effects are rather small as long as 
the angular aperture is kept sufficiently small. Here we will examine more
closely focusing of atoms with a tightly focused radial laser beam. The 
advantage of such a scheme is that the beam is symmetric, and only the z 
component is significant when the angular aperture is large and the 
center beam is blocked out.

The radial polarization distribution is perhaps better suited to provide the
necessary symmetry. In fact, we have seen that by using a large aperture 
combined with a suitable aperture, only the z component will be of importance. 
To that end, the aberration function is given by   
\begin{equation}
\Phi(\rho) = \frac{\pi \rho ^{2}}{\lambda _{D} f}- A J_{0} ^{2} (k\rho ) \,\,\, ,
\label{ab}
\end{equation}
where A is the socalled field area,
\begin{equation}
A=\frac{\tau k^{2} C_{1}^{2} \Gamma ^{2} }{16\eta I_{s} \Delta } \,\,\, .
\end{equation}
The field area is easily modulated by changing the power of the laser beam.
Here we are mostly interested in a general description and will not be
concerned with the exact laser power, which should be optimized in
the particular experimental setup.
 
\begin{figure}
\includegraphics[width=12cm]{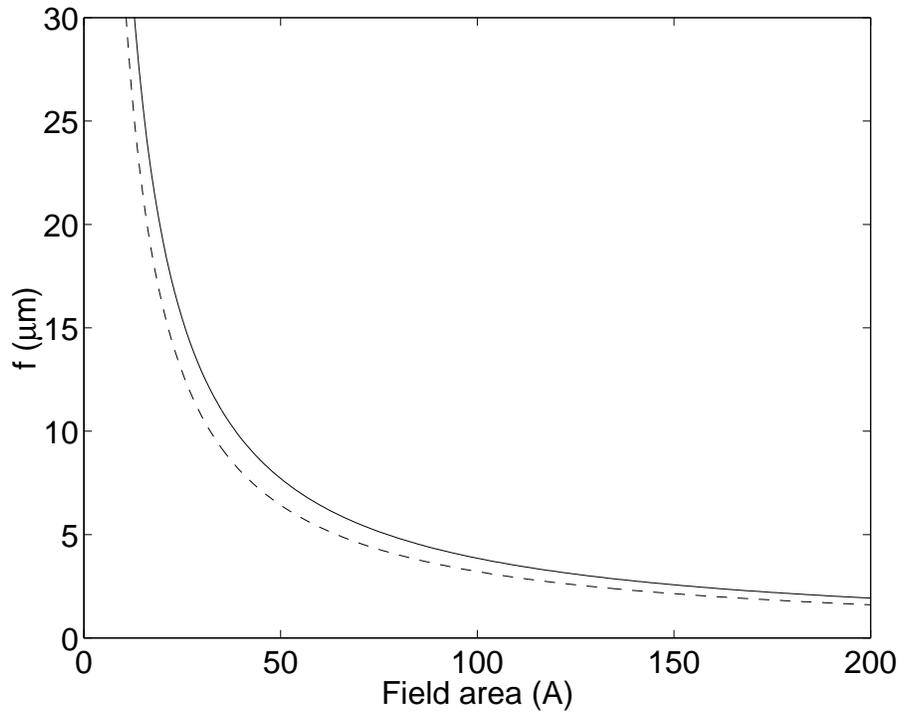}
\caption{\label{f5} The focal distance for focusing of $^{52}$Cr (solid line) and 
$^{23}$Na (dashed line). }
\vspace{2cm}
\end{figure}

\begin{figure}
\includegraphics[width=12cm]{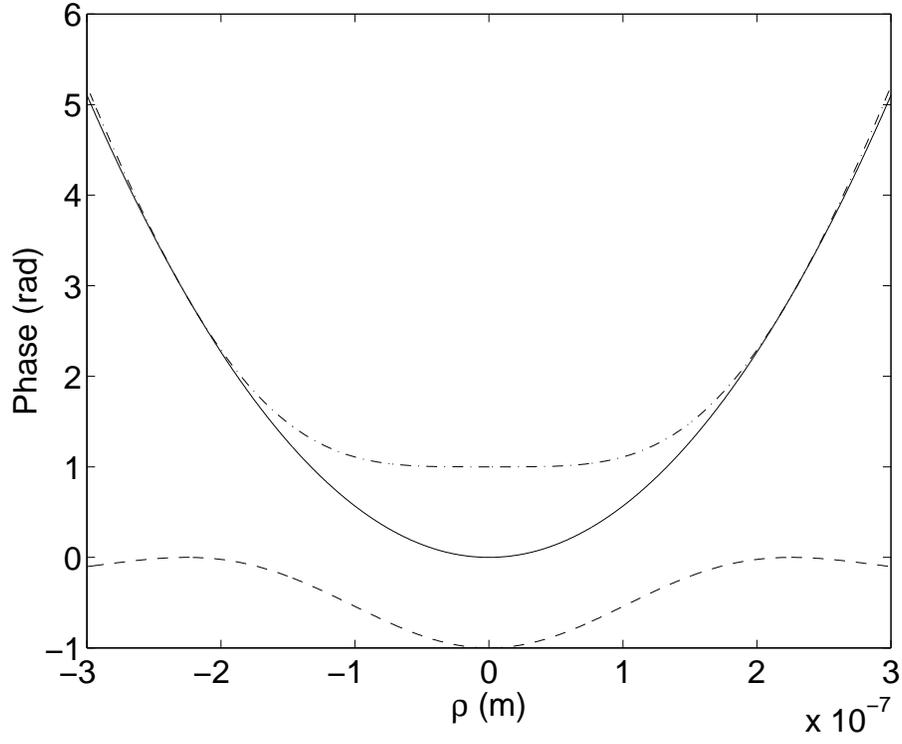}
\caption{\label{f6} The wavefront aberrations when A=-1. The solid line shows 
$\pi \rho ^{2}/\lambda _{D} f$, the dashed line $-A J_{0} ^{2} (k\rho )$ and the 
dash-dotted line $\Phi(\rho)$. Note that only in a small area around the optical
focal point the phase function is constant (which means that the aberrations are
small).}
\vspace{2cm}
\end{figure}

\begin{figure}
\includegraphics[width=12cm]{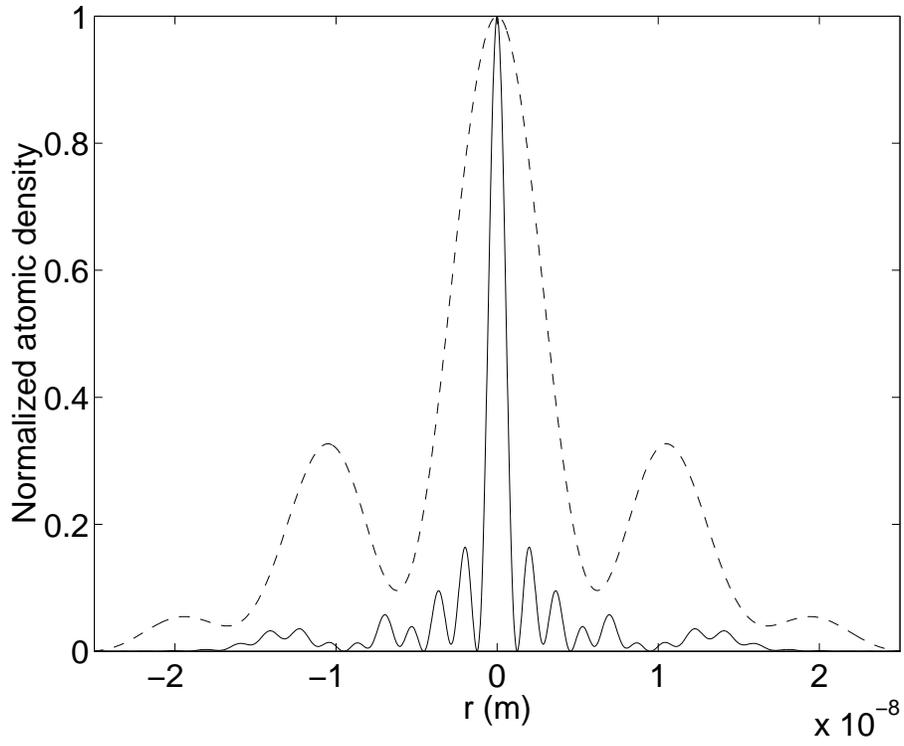}
\caption{\label{f7} The atomic density when A=-100 (solid line) and A=-20 (dashed line). }
\vspace{2cm}
\end{figure}

For small values of $\rho$, the focal distance in the laboratory frame is
\begin{equation}
f^{r}=\frac{\lambda ^{2}}{2\pi A \lambda _{D}} \,\,\, . 
\end{equation}
In Fig. \ref{f5} $f^{r}$ is plotted as a function of the field area for two 
different atoms with the same intial velocity (100 m/s). The solid line
corresponds to $^{52}$Cr with $m=8.68 \times 10^{-26}$ kg and 
$\lambda$ = 0.43 $\mu m$, whereas the dashed line 
corresponds to $^{23}$Na with $m=3.84 \times 10^{-26}$ kg and $\lambda$ = 0.59
$\mu m$. One sees that typical focal distances range from a few microns to
hundred microns. However, care must be taken when selecting the field area,
since this together with the aperture determine the aberrations of the
focusing system. Figure \ref{f6} shows the phase function (dash-dotted line) for Cr atoms with 
velocity 500 m/s and A=-1. It is seen
that the phase function is nearly constant within an aperture of $\sim$ 0.1$\mu m$, 
where the aberrations are negligible. Figure \ref{f7} shows $|\Psi (r)|^{2}$ for focusing of $^{52}$Cr with a 
velocity of 500 m/s and aperture a = 0.1 $\mu m$. The solid line shows the 
atomic density when A = -100 (A is negative because
the laser light is red-detuned), whereas the dashed line corresponds to A=-20. 
Here the focal lengths are 19 and 96 $\mu m$, respectively.
We used a rather narrow atomic beam in order to block out most of the 
aberrations, thus allowing a resolution of $\sim$ 1 nm (A=-100).

\subsection{Blue-detuned focusing}
In the case of blue-detuned focusing, the atom-field detuning is large and
positive, $\Delta \gg \Gamma$. Thus, the atom will be expelled from the high
intensity region. For this purpose the intensity near the focal region of an 
azimuthal polarization distribution aberration may be useful. In this case the
aberration function is
\begin{equation}
\Phi(\rho) = \frac{\pi \rho ^{2}}{\lambda _{D} f}- A J_{1} ^{2} (k\rho ) \,\,\, ,
\label{aa}
\end{equation}        
which is identical to that used by Dubetsky and Berman\cite{Dubetsky}, except for the constant
A. This is due to the fact that we consider here only the
intensity gradient and neglect any effects of the polarization gradient. 
For small values of $\rho$, the focal distance in the laboratory frame is
\begin{equation}
f^{a}=\frac{\lambda ^{2}}{\pi A \lambda _{D}} \,\,\, . 
\end{equation} 

\section{Discussion}
\begin{figure}
\includegraphics[width=12cm]{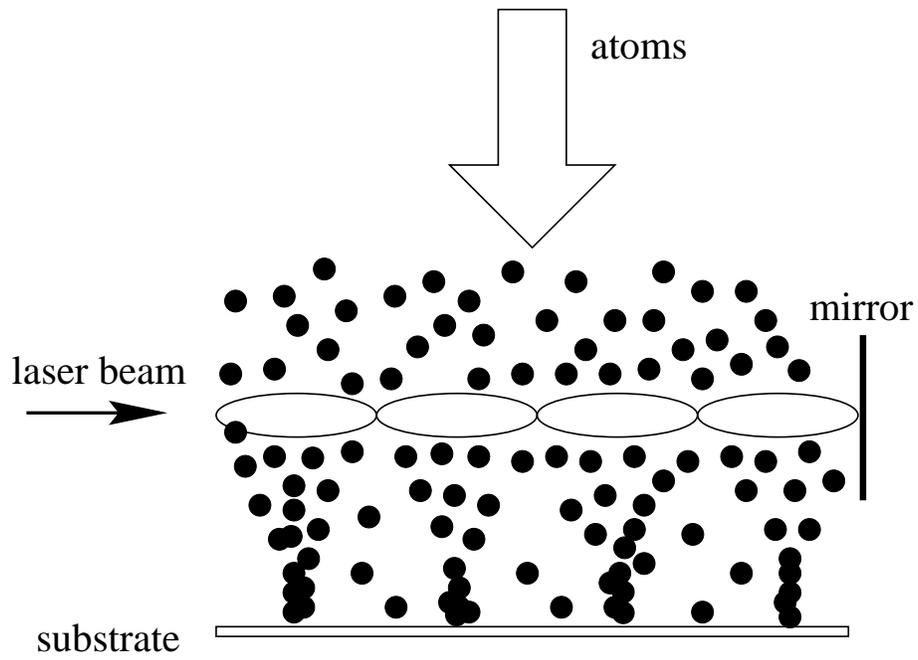}
\caption{\label{f8} Deposition of atoms on a substrate using a standing wave. }
\vspace{2cm}
\end{figure}

\begin{figure}
\includegraphics[width=12cm]{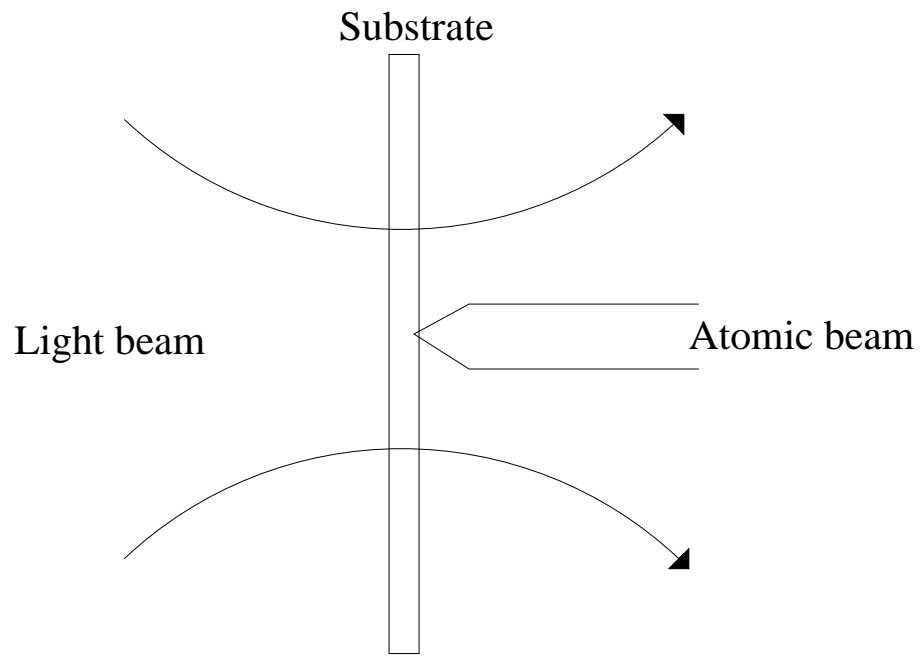}
\caption{\label{f9} Deposition of atoms on a transparent substrate with a focused light
beam.}
\vspace{2cm}
\end{figure}

Two systems for practical atom lithography are shown in Figs. \ref{f8} and
\ref{f9}, respectively. The first, which applies a 1D standing light wave, has 
been successfully adopted for focusing of many different atoms\cite{Rehse}. The
disadvantage of this system is the low power and gradients (since the light is distributed
over a large area). A more
promising scheme for depositing atoms directly onto a transparent substrate 
is shown in Fig. \ref{f9}. Here the atoms are incident on the opposite side of
the substrate, and the lens system do not disturb the atomic beam. In practice
the light beam should be slightly defocused on the far side of the substrate
(where the atoms come in). 
Thus, if the light beam is properly corrected for the spherical aberrations
introduced by the substrate, the electric field is the same as in Sec. II.     
We have shown that such a system could in principle give a resolution of 1 nm,
but the system requirements are rather strict. Thus, it is indeed necessary to
perform further investigations before a scheme similar to that of Fig. \ref{f9}
can be applied in practice.   

It has already been mentioned that the spontaneous emission aberrations must be
reduced to a minimum in order to achieve a diffraction limited distribution. 
This can be obtained by using a pulsed laser beam, typically of pulse length 
10 ns or shorter. Moreover, it is also of importance to reduce the chromatic aberrations. 
Again, a pulsed laser seems to be the solution,
since then the focusing properties are less dependent on the atomic
velocity. On the other hand, it should be pointed out that for an atom laser
(where the atoms are monochromatic) chromatic aberrations are not a problem.

Notice that both for radial and
azimuthal beams it is possible to alter the intensity gradients by manipulating
the phase and amplitude distribution at the exit pupil. A more general 
formulation of the electromagnetic field in presence of arbitrary exit pupils
are found in Ref. \cite{Helseth}. By inserting an appropriate aperture it 
should be possible to manipulate the optical intensity gradient so that it looks 
more like a parabola. A further improvement would be to use an annular atomic beam. In this way one may use a smaller part
of the optical potential. Then one may expect smaller aberration, since it 
should in principle be easier to make a small part of the optical intensity
distribution look like a parabola than the whole atomic aperture. Naturally, this approach would lead to
lower atomic densities, which in many applications is a big disadvantage.

\section{Conclusion}
Focusing of inhomogenously polarized light is studied, and expressions for the
electric field in the focal region are given. It is seen that some 
polarization distributions at the exit pupil may give intensity distributions 
near the focal plane which are of interest for focusing of atoms. To this end, 
focusing of an atomic beam using near-resonant laser light is also studied, 
and the wave function in the focal region is found. Moreover, it is 
suggested how different light forces may be used in red and blue-detuned 
focusing of atoms.

\newpage

\newpage

\newpage
  
\newpage
  
\newpage

\newpage

\newpage

\newpage

\newpage

\end{document}